\documentclass[%
 twocolumn,
 amsmath,amssymb,
 aps,
]{revtex4-1}

\usepackage{graphicx}% Include figure files
\usepackage{dcolumn}% Align table columns on decimal point
\usepackage{bm}% bold math
\usepackage{float}
\begin{document}

\newcommand\needref{{${}^{\it\bf ref\quad}$}}
%\preprint{APS/123-QED}

\title{Pseudo-Magnetic  Quantum Hall Effect In Oscillating Graphene}% Force line breaks with \\

\author{Anita Bhagat}
 \email{anita@ou.edu}%{Physics Department, The University of Oklahoma.}%Lines break automatically or can be forced with \\
\author{Kieran Mullen}% 
 \email{kieran@ou.edu}
\affiliation{%
Homer L. Dodge Dept. of Physics and Astronomy\\
 The University of Oklahoma 
}%

\date{\today}% It is always \today, today,
\begin{abstract}
When a graphene layer is stressed, the strain alters the phase an electron accumulates hopping between sites in a way that can be modeled as arising from a pseudo-magnetic vector potential. We examine the case of an oscillating graphene ribbon and explore a new effect - an oscillating resistance  arising from an oscillating quantum Hall effect. This pseudo-magneto-resistance is large, and depends upon the frequency and the amplitude of the acoustic oscillations. We calculate the consequences for experiment. 
%\begin{description}
%\end{description}
\end{abstract}

\maketitle

%\tableofcontents

\def\blah{{\bf blah, blah, blah}}

\section{Introduction}

In 2004 Geim {\it et al.} isolated a single layer of graphite, setting off the discovery of a long list of remarkable properties of graphene, such as its great strength, high mobility, and linear electronic energy spectrum.\cite{GrapheneReview1,GrapheneReview2}
   One additional extraordinary property it displays is a strain-induced pseudo-vector potential.\cite{pseudoB1} The geometrical deformation of the monolayer graphene lattice introduces strain which can alter  the phase difference between adjacent sites in a tight-binding model of the system. This phase difference
   can be viewed as arising from a ``pseudo-magnetic field''.\cite{vectorpotential2}  
   Such a pseudo-magnetic field does not break time reversal symmetry because it couples with opposite sign to electrons occupying different Dirac points in the band structure. One consequence of this was dramatically
   demonstrated by the observation of quantized Landau levels in strained graphene in the absence of an external magnetic field.\cite{qHallExpt1} This effect can be substantial: strain-induced vector potentials in static monolayer graphene have produced pseudo magnetic fields of 300~T.\cite{qHallExpt1}

There has been much theoretical and experimental work done to study the pseudo-magnetic field created by  strain applied  to a static graphene layer.\cite{pseudoB3, pseudoB4, pseudoB5, pseudoB6, pseudoB7, pseudoB8, pseudoB9, pseudoB10, qHallExpt1,qHallExpt2,qHallExpt3,vectorpotential1,vectorpotential2, scalarPotential3} In this paper we will discuss the physics of a rapidly oscillating pseudo-magnetic field generated by an acoustically driven graphene ribbon. Typical electronic relaxation times in graphene are on the order of picoseconds.\cite{eeScatt, relax} If models developed for static lattice distortions are valid for the relatively ``slow'' acoustic oscillations, the system can potentially have pseudo-magnetic fields of several Tesla oscillating at a kilohertz.  Such a regime is inaccessible for normal magnetic fields and opens up many new possibilities.   

In this work we start with a review of the connection between pseudo-magnetic fields and  applied strain. We then investigate oscillating pseudo-quantum Hall effect phenomena for low density graphene nano-ribbons with simulated experimental results for realistic parameters. The relevant experimental frequencies are quite low - indeed the adiabatic approximation gets better the lower the frequency is. For fixed frequency, the magnitude of the affect depends on the amplitude of the distortion giving a separate control for experiment. This independent control 
may allow investigation above and below the adiabatic limit. In this work we ignore the pseudo-electric field induced by strain. While in-plane electric fields can
cause a collapse of Landau levels in graphene \cite{relax1,relax2}, the in-plane fields generated in this paper are small, and by appropriate choice of polarization can be minimized.

\begin{figure} [b]
\includegraphics[width=2.7in]{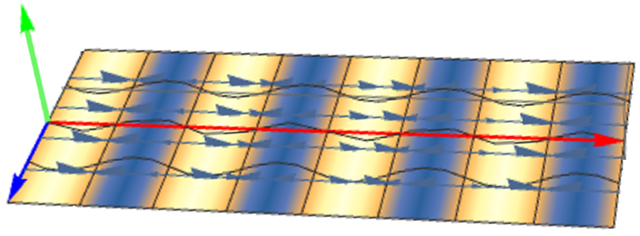}

 \caption{A schematic representation of displacement in a suspended graphene nano-ribbon driven acoustically to create an oscillating pseudo-magnetic field, with the blue, red and green arrows representing the x, y and z directions, respectively. The displacement is in the y-direction, producing a vector potential in x and a magnetic field in the z direction. The amplitude of the magnetic field in the z-direction is denoted by the shading, with large positive fields shaded in yellow.
 % The  wavelength is 4\,$ {\mu }$m.
   \label{fig:expt}}  
 \end{figure}

\subsection{Basic Theory}

 We consider an ideal 2D graphene nano-ribbon suspended between supports. The ribbon is driven acoustically by a piezoelectric transducer to produce a longitudinal standing wave. An atom at the position \,\,\, \,\,\,\,\,\,                   \,\,\,\,\,              
$\vec r \equiv x \hat i + y \hat j + z \hat k$ is displaced by vector $ \vec {u} {(\vec r,t)} $  (Fig.\ref{fig:expt})

\begin{equation}
 \vec {u} {(\vec r,t)} =   u_0 \sin{(k_y y)}  \cos{(\omega t)} \,\, {\hat j}       \label{eqnDisplacement}                                          
\end{equation}

 \begin{figure}[h]
\includegraphics[width=3.0in]{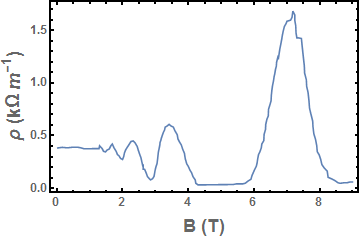}
 \caption{Quantum Hall effect in 2D graphene nano-ribbon. The longitudinal magneto-resistivity in 2D graphene is plotted as a function of real magnetic field, reproduced with permission from \cite{janssen}. \label{fig:janssen}}
 \end{figure}
 
 producing a standing wave with displacements in the y direction. The stretching in the graphene layer produced by displacement produces a strain tensor $u_{ij}(\vec r,t)$ which is given by
\begin{equation}
 \ u_ {xx} = \frac {\partial u_x}{\partial x};\qquad \ u_ {yy} = \frac {\partial u_y}{\partial y}; \qquad \ u_ {xy} = \frac{1}{2}({\frac {\partial u_x}{\partial y}+\frac {\partial u_y}{\partial x}})                                            
\end{equation}

This strain tensor alters the phase difference between atoms in a tight-binding model of graphene in a fashion 
equivalent to one induced by a gauge field $ \vec {A_s} $ (pseudo-magnetic vector potential).\cite{pseudoB2,pseudoB3}

\begin{equation}
\vec {A_s} = \displaystyle{\frac {c \,{\hbar}\,{ \beta\tau}} {e \,a_o}\left[{\begin{matrix} u_{xx} - u_{yy} \\ -2u_{xy} \end{matrix}} \right ]} 
\end{equation}

where ${a_0 \approx}$ 1.4 A${^o}$ is the length of the bond between neighboring carbon atoms and $\beta$ and c are dimensionless parameters that  are equal to 2 and 1 respectively.\cite{pseudoB1}
The quantity $\tau$ is the valley pseudo-spin  (called ``valley spin'' hereafter to avoid confusion with the layer index in multi-layer systems, which is 
also called ``pseudo-spin'') , taken as $+1$ at the $K$ point in the 2D bandstructure, and $-1$ at the $K'$ point.  
Therefore, even in the absence of a real magnetic field a strong pseudo-field is observed which comes from the curl of  $ \vec {A_s} $, but it is of opposite sign for electrons of opposite ``valley-spin''.  For the displacement of eq.(\ref{eqnDisplacement}) the magnetic field is given by 

\begin{equation}
\vec B_s(\vec r)= \frac {- {\hbar}\,{ \beta \tau  \, u_0\, k_y{^2} }}  {e \,a_0}\,   {\sin{(k_y y)} \cos{(\omega t})}\,{\hat k}
\end{equation}

 We have introduced a dimensionless strain amplitude  \,f = $ \frac{\bigtriangleup L}  {L}$. The strain amplitude is expressed as the ratio of the wave amplitude to a half wavelength, so that in terms of wave number,   $f=\frac{u_0 k} {\pi } $.

\subsection{Oscillating quantum Hall magneto-resistance}

We assume a suspended graphene nano-ribbon set up for a two-terminal measurement, driven with an acoustic oscillation of a wavelength on the order of several microns. The oscillation has a wavelength much longer than the coherence length of electrons in the system, and a frequency far less than the characteristic electron-electron scattering rate i.e. on the order of $10^{-11}$ $s^{-1}$.\cite{eeScatt}
 \begin{figure}[b]
 \centerline{\includegraphics[width=3.0in]{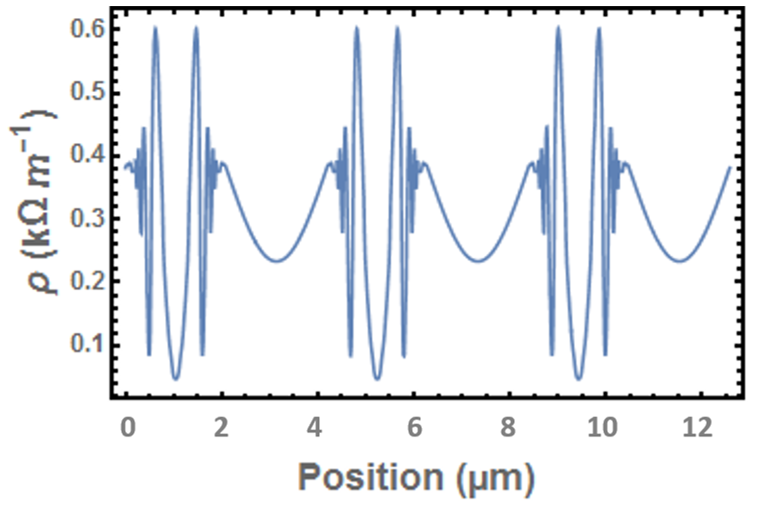}}
\centerline{\large(a)}
\centerline{\includegraphics[width=3.0in]{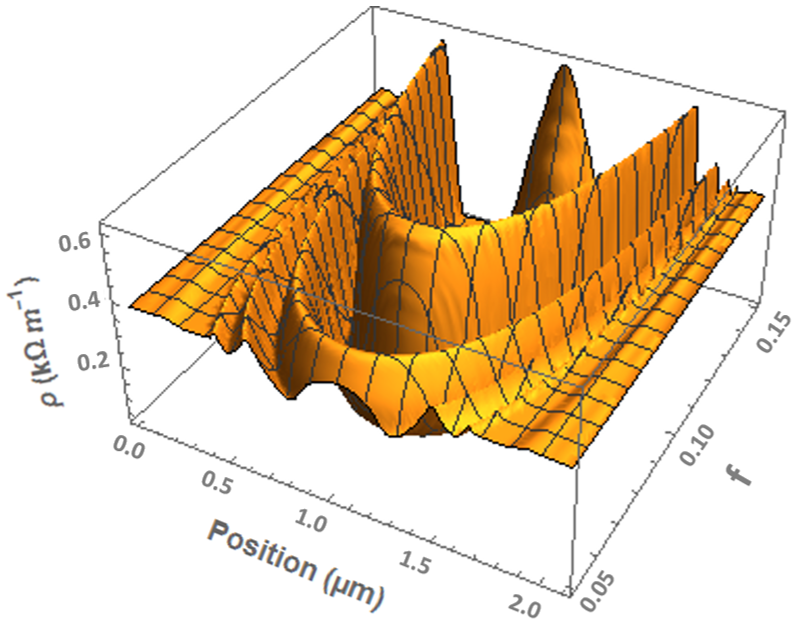}}
\centerline{\large(b)}
%\Shortstack{{\littleA} (a)  {\bigA} (b)}
%\Shortstack{{\littleA} (a) } \Shortstack{{ \bigA} (b)  }

% \LittleA
 %\LittleB
% \Shortstack{{\LittleA}  { (a)}   { \LittleB} (b) } 
  \caption{Longitudinal resistivity of an oscillating graphene nano-ribbon. {(a)} Resistivity as a function of position along the length of the monolayer graphene nano-ribbon for strain amplitude $ f = 0.1 $, and wavelength  $\lambda$ = 4\,$ {\mu }$m  at a time $\omega t= 2 \pi n$, for a sample assumed to show the longitudinal resistance of Fig.(\ref{fig:janssen}). {(b)} Plot of local resistivity as a function of strain and position for $\vec B_s(\vec  u(\vec r))$ using interpolation of Fig.(\ref{fig:janssen}). The pseudo-magnetic field rises and than falls  along the x-direction of the oscillating graphene nano-ribbon
  %  for resistivity of around 2 $ K \Omega m^{-1} $ 
as seen in figure (a). The 3D plot shows this variation in the  pseudo-magnetic field with the increase in strain amplitude $f$. When $f$ is increased, more structures are visible, corresponding to the peaks at higher field in figure(a). \label{fig:resistivity1}}
 \end{figure}
Since the acoustic oscllations of the ribbon are slow compared to the motion of the electrons, we treat each portion of the ribbon with displacement $\vec u(\vec r)$ as an independent, equilibrium, quantum Hall ``sample'' with a magnetic field that is given by the pseudo-magnetic field of eqn.(4). The oscillating ribbon has pseudo-magnetic field that varies as a function of space as the effective strain amplitude, $f \sin{(k_y y)}$, varies in space. To obtain the longitudinal resistance $\rho_{xx}$ for any positive or negative value of this oscillating pseudo-magnetic field, we use the value it would have in a {\it uniform}  magnetic field $\vec B_s(\vec u( \vec r))$ using Fig.(\ref{fig:janssen}). Since the longitudinal resistance in quantum Hall systems is non-universal, we take a representative value from published experimental  data \cite{janssen} on graphene in Fig.(\ref{fig:janssen}). The oscillating strain produces a pseudo-magnetic field standing wave
 with a maximum amplitude that potentially can reach several Tesla. The resistivity obtained from the interpolation of Fig.(\ref{fig:janssen}) will therefore also  oscillate  in space and time,  % with a maximum value fluctuating around 2 $ K \Omega m^{-1} $ 
 (Fig.\ref{fig:resistivity1}a). %  for a wavelength 4\,$ {\mu }$m. 
 The oscillations of pseudo-field  display more structure as the strain amplitude is increased, as shown in Fig.(\ref{fig:resistivity1}b)

  \begin{figure}[t]
\centerline{\includegraphics[width=3.2in]{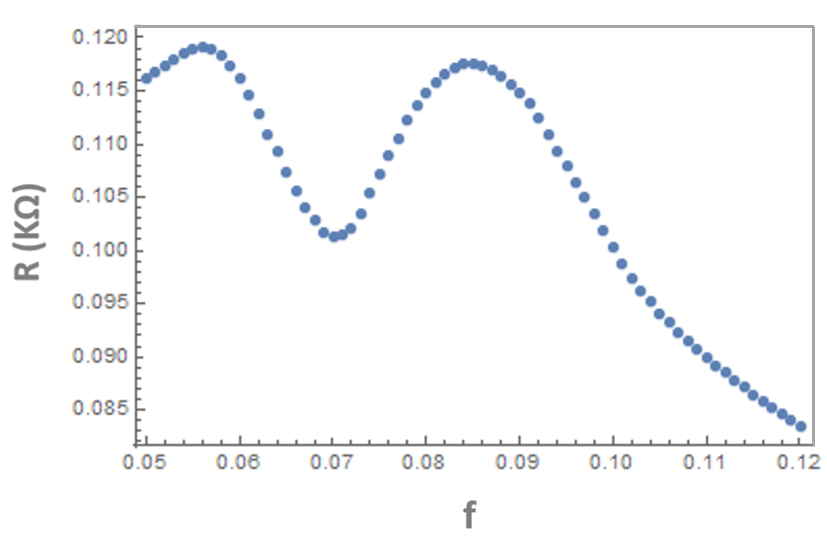}}

 \caption{(a)Plot of resistance amplitude at times when the oscillation  is at a  maximum, $\omega t=2 \pi n$, as a function of 
 strain amplitude $f$. The total resistance is calculated by integrating the resistivity of Fig.(\ref{fig:resistivity1}) across one wavelength. The wavelength of the oscillation is $\lambda = \rm 4.0\mu m$, and the resistance amplitude is calculated using the experimental data of ref.(\cite{janssen}).} \label{fig:resistivity2}
 \end{figure}
 
   \begin{figure}[b]
\centerline{\includegraphics[width=3.0in]{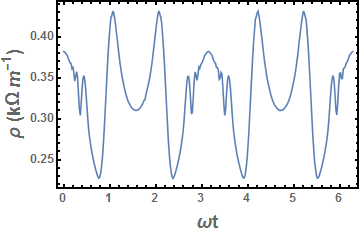}}

 \caption{Plot of the spatially averaged  resistance of a nano-ribbon  as a function of time for  a dimensionless strain amplitude,  $f=\frac{u_0 k} {\pi }=0.10$, where $u_0$ is the strain amplitude of a standing wave of wavenumber $k$.  The resistance oscillates as a function of time because the effective strain amplitude, $f \cos{(\omega t)}$, oscillates in time. The resistance at any instant in time corresponds to integrating over a position slice
 of the resistivity plotted in  Fig.(\ref{fig:resistivity1}b)  for some value of this effective strain amplitude. } \label{fig:rhoVtime}
 \end{figure}

 It is important to note that this effect does not break time reversal symmetry. Electrons from different K-points see opposite signs of the pseudo-magnetic field, but the magneto-resistance is independent of the sign of the field, so both valley-spins polarizations are gapped (or not gapped) at exactly the same value of the strain.

 %\begin{figure}[b]
 %\includegraphics[width=3.5in]{Resisitivity.png}
 %\includegraphics[width=3.5in]{Resisitivity1.png}
 %\caption{Longitudinal resistivity of an oscillating graphene nano-ribbon. {(a)} Resistivity as a function of position along the length of the monolayer graphene nano-ribbon for u = 0.1  strain amplitude and $\lambda$ = 4\,$ {\mu }$m wavelength for a sample assumed to show the longitudinal resistance of Fig.\ref{fig:janssen}. {(b)} Plot of local resistivity as a function of strain and position for $\vec B_s(\vec  u(\vec r))$ using interpolation of  Fig.\ref{fig:janssen}. The pseudo-magnetic field rises and than falls  along the x-direction of the oscillating graphene nano-ribbon for resistivity of around 2 $ K \Omega m^{-1} $ as seen in figure (a) . The 3D plot shows this variation in the  pseudo-magnetic field with the increase in strain amplitude ``f''. When f is increased, more structures are visible, corresponding to the peaks at higher field in figure(a). \label{fig:resistivity1}}
 %\end{figure}
 
A more experimentally accessible property for a mechanically oscillating ribbon is its total resistance. For the oscillating monolayer graphene the total resistance at any point in time will be given by numerically integrating the resistivity graph of Fig.(\ref{fig:resistivity1}) along the length of the ribbon. If we choose a specific point in the oscillation, say when $\omega t = 2 \pi n$, we can calculate the expected resistance at that point as a function of strain
amplitude. This  calculated resistance  varies non-monotonically  as a function of strain amplitude as shown in Fig.(\ref{fig:resistivity2}). It oscillates because  increasing the strain amplitude may allow the pseudo-magneto-resistance to reach a higher resistance peak in Fig.(2), increasing the total resistivity, but 
increasing the strain further brings it into a regime of lower resistivity while simultaneously reducing the width of the higher resistance region. The width is reduced because the entire range of the sweep of the pseudo-magnetic field must still fit within one wavelength, and increasing the strain amplitude does not change the wavelength of the oscillation. The lower minima in the oscillations at larger strain are a reflection of the fact that the experimental data for $\rho_{xx}$ has deeper minima at larger fields.   
 
  Alternatively, we may plot the resistance of the ribbon as a function of time as the standing wave goes through one oscillation. If Fig.(\ref{fig:resistivity1}a) represents a slice of Fig.(\ref{fig:resistivity1}b) at constant strain amplitude $f$, then the sinusoidal time dependence of the oscillation corresponds in effect to sweeping this slice from $f=0$ to the some maximum value and back. To get the resistance at any time $t$ we simply must integrate the corresponding slice along the $x$ direction. An example of a resistance trace as a function of time for a given strain amplitude is given in Fig.(\ref{fig:rhoVtime})
 
  \begin{figure}[h]

\includegraphics[width=3.0in]{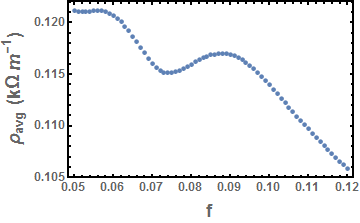}
 \caption{ Plot of time and spatially averaged resistance of a nano-ribbon  as a function of dimensionless strain amplitude,  $f=\frac{u_0 k} {\pi } $ where $u_0$ is the strain amplitude of a standing wave of wavenumber $k$.} \label{fig:rhoavg}
 \end{figure}

Finally, we may plot the time average of the resistance of the ribbon as a function of strain amplitude, as shown in Fig.(\ref{fig:rhoavg}). As can be seen from above, much detail of the structure will be lost by integrating the resistivity  over both space and time. However, this is the simplest and most straightforward measurement that would show this fundamental quantum mechanical effect from an acoustic oscillation.

\subsection{Conclusion}

Experiments have already verified that a static strain in graphene can produce a pseudo-magnetic field of many Tesla.  
If the static distortion calculation is valid for  the ``slow'' distortion of an acoustic wave, then pseudo-magnetic fields of several Tesla could be observed to oscillate at high frequency, a previously inaccessible regime of  electron dynamics.
We have investigated an oscillation in the resistance due to a quantum Hall-like effect  produced from time dependent pseudo-magnetic field. This phenomena should be observable at experimentally accessible frequencies and temperatures. The absence of this phenomenon is equally intriguing since  then the cross-over from the experimentally validated static theory to a time dependent distortion must itself be investigated.  
   
\bigskip

\,{\it Acknowledgements.}$-$ K. M. was supported by NSF grant DMR-1310407. \\

\end{document}